\documentclass[prd,twocolumn,a4paper,showpacs,superscriptaddress,floatfix]{revtex4}

\usepackage{pslatex}
\usepackage{graphicx}
\usepackage{amsmath}

\newcommand{\be}{\begin{equation}}
\newcommand{\ee}{\end{equation}}

\newcommand{\reff}[1]{(\ref{#1})}

\begin{document}

\title{QCD with two flavors of Wilson fermions:\\The QCD vacuum, the Aoki vacuum 
 and other vacua}

\date{\today}
\author{V.~Azcoiti}
%\email{azcoiti@azcoiti.unizar.es}
\affiliation{Departamento de F\'{\i}sica Te\'orica, Facultad de Ciencias, 
Universidad de Zaragoza,
Cl. Pedro Cerbuna 12, E-50009 Zaragoza (Spain)}
\author{G.~Di Carlo}
%\email{gdicarlo@lngs.infn.it}
\affiliation{INFN, Laboratori Nazionali del Gran Sasso, I-67010 Assergi, 
(L'Aquila) (Italy)}
\author{A.~Vaquero}
%\email{alexv@unizar.es}
\affiliation{Departamento de F\'{\i}sica Te\'orica, Facultad de Ciencias, 
Universidad de Zaragoza,
Cl. Pedro Cerbuna 12, E-50009 Zaragoza (Spain)}

\begin{abstract}
We discuss the vacuum structure of QCD with two flavors 
of Wilson fermions. We derive two possible scenarios: (i) If the spectral 
density $\rho_U(\lambda,\kappa)$ of the overlap hamiltonian in a fixed 
background gauge field is not symmetric in $\lambda$, Hermiticity is violated 
and Hermiticity violation effects 
could influence numerical determinations of the $\eta$ meson mass if we are 
not near enough to the continuum limit, where Hermiticity should be recovered; 
(ii) otherwise we argue that, under certain assumptions, new phases appear beside the Aoki phase, which can be 
characterized by a nonvanishing vacuum expectation value of 
$i\bar\psi_u\gamma_5\psi_u+i\bar\psi_d\gamma_5\psi_d$, 
and with vacuum states that cannot be connected with the Aoki vacua by parity-flavor symmetry 
transformations. Quenched numerical simulations suggest that the second 
scenario is more likely realized.
\end{abstract}
\maketitle

\section{Introduction}

Since the first numerical investigations of four-dimensional 
non-abelian gauge theories 
with dynamical Wilson fermions were performed in the early 80's 
\cite{an,hamber}, 
the understanding of the phase and vacuum structure of lattice QCD with 
Wilson fermions at non-zero lattice spacing, and of the way in which chiral 
symmetry is recovered in the continuum limit, has been a goal of lattice 
field theorists. The complexity of the phase structure of 
this model was known a long time ago. The existence of a phase with parity 
and flavor symmetry breaking was conjectured for this model by Aoki in the 
middle 80's \cite{a1,a2}. From that time on, much work has been done in order 
to confirm this conjecture, to establish a quantitative phase diagram 
for lattice QCD with Wilson fermions, and to delimit the parameters region 
where numerical calculations of physical quantities should be performed.
References \cite{a1}-\cite{ilg3} are a representative but incomplete list 
of the work done on this subject.

In this paper we analyze the vacuum structure of lattice QCD with
Wilson fermions at non-zero lattice spacing, with the help of the probability 
distribution function of the parity-flavor fermion bilinear order parameters 
\cite{pdc}\cite{vw2}. We find that if the spectral
density $\rho_U(\lambda,\kappa)$ of the overlap hamiltonian, or Hermitian 
Dirac-Wilson operator, in a fixed
background gauge field $U$ is not symmetric in $\lambda$ in the 
thermodynamical limit for the relevant gauge configurations, Hermiticity of 
$i\bar\psi_u\gamma_5\psi_u+i\bar\psi_d\gamma_5\psi_d$ will 
be violated at finite $\beta$. Assuming that the Aoki phase ends at finite 
$\beta$, this would imply the lost of any physical interpretation of this 
phase in terms of particle excitations. In addition, the lost of Hermiticity 
of $i\bar\psi_u\gamma_5\psi_u+i\bar\psi_d\gamma_5\psi_d$ at finite $\beta$ 
also suggests that a reliable determination of the $\eta$ mass would require 
to be near enough the continuum limit. If on the contrary the spectral
density $\rho_U(\lambda,\kappa)$ of the Hermitian Dirac-Wilson operator
in a fixed background gauge field $U$, is symmetric in $\lambda$, we explain how, under certain assumptions, the 
existence of the Aoki phase implies also the appearance of other phases, in 
the same parameters region, which can be characterized by a non-vanishing vacuum expectation value of
$i\bar\psi_u\gamma_5\psi_u+i\bar\psi_d\gamma_5\psi_d$,
and with vacuum states that can not be connected with the Aoki vacua by parity-flavor symmetry
transformations.

The outline of this paper is as follows. In Sec. II we derive the $p.d.f.$
of $i\bar\psi\gamma_5\tau_3\psi$ and $i\bar\psi\gamma_5\psi$ and analyze the 
conditions to have an Aoki phase with spontaneous parity-flavor symmetry 
breaking. Section III contains our derivation of the p.d.f. of the same 
fermion bilinears of Sec. II but in presence of a twisted mass term in 
the action. We discuss also in this Sec. how if the spectral
density $\rho_U(\lambda,\kappa)$ of the Hermitian Dirac-Wilson operator
in a fixed background gauge field $U$ is not symmetric in $\lambda$, 
a violation of Hermiticity manifests in a negative vacuum expectation value 
for the square of the Hermitian operator $i\bar\psi\gamma_5\psi$ in the 
infinite lattice limit. In Sec. IV we discuss the other possible scenario 
i.e., we assume a symmetric spectral density $\rho_U(\lambda,\kappa)$ and 
then derive, assuming that the Aoki vacuum does exists,  that new vacuum 
states should appear. These new vacua, which are not
connected with the Aoki vacua by parity-flavor symmetry
transformations, are analyzed in Sec. V. 
Section VI contains some numerical results obtained by diagonalizing  
quenched configurations in $4^4, 6^4, 8^4$ lattices. The goal of this 
analysis was to distinguish between the two possible scenarios. Section 
VII contains our conclusions.

\section{The Probability Distribution Function in the Gibbs State}

The model we are interested in is QCD with two degenerate Wilson quarks. 
The fermionic part of the Euclidean action is 

\be
S_F=\bar\psi W(\kappa) \psi
\label{acfer}
\ee

\noindent
where $W(\kappa)$ is the Dirac-Wilson operator, $\kappa$ is the hopping 
parameter, which is 
related to the bare fermion mass $m_0$ by $\kappa = 1/(8+2m_0)$, and 
flavor indices are implicit ($W(\kappa)$ is a two-block diagonal matrix).
The standard wisdom on the phase diagram of this model in the gauge 
coupling $\beta, \kappa$ plane is the one shown in Fig. 1. The two different 
regions observed in this phase diagram, A and B, can be characterized as 
follows; in region A parity and flavor symmetries are realized in the vacuum, 
which is supposed to be unique. The Gibbs state is then very simple in this 
region, and continuum QCD should be obtained by taking the $g^2\rightarrow 0$, 
$\kappa\rightarrow 1/8$ limit from within region A. We will call region A 
the QCD region. In region B, on the contrary, parity and flavor symmetries are 
spontaneously broken, there are many degenerate vacua connected by 
parity-flavor transformations in this region, and the Gibbs state is 
therefore made up from many equilibrium states. In what follows we will call 
region B as the Aoki region.

\begin{figure}[h]
\resizebox{8 cm}{!}{\includegraphics{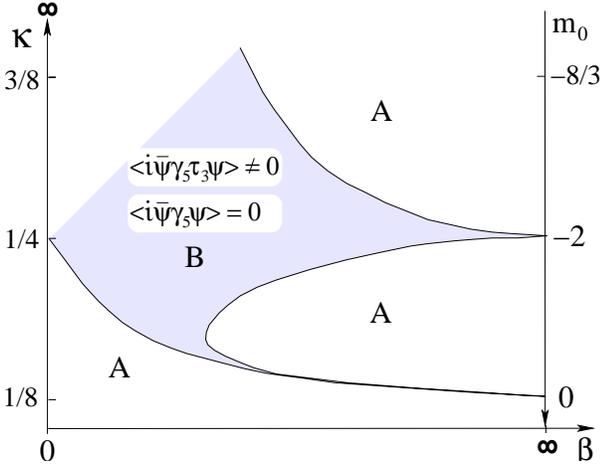}}
\caption{Aoki (B) and physical (A) region in the $(\beta,\kappa)$ plane. Adapted from \cite{ilg} with courtesy of the authors.}
\end{figure}

The order parameters to distinguish the Aoki region from the QCD region are 
$i\bar\psi\gamma_5\tau_j\psi$ and $i\bar\psi\gamma_5\psi$, with $\tau_j$ 
the three Pauli matrices. For the more standard election, $j=3$, they 
can be written in function of the up and down quark fields as follows

$$
i\bar\psi\gamma_5\tau_3\psi=i\bar\psi_u\gamma_5\psi_u-i\bar\psi_d\gamma_5\psi_d
$$
\be
i\bar\psi\gamma_5\psi=i\bar\psi_u\gamma_5\psi_u+i\bar\psi_d\gamma_5\psi_d
\label{orpa}
\ee

The Aoki phases are characterized by \cite{a2}

$$
\langle i\bar\psi\gamma_5\tau_3\psi \rangle \neq 0
$$
\be
\langle i\bar\psi\gamma_5\psi \rangle = 0
\label{orpaao}
\ee

The first of the two condensates breaks both parity and flavor symmetries. 
The non-vanishing vacuum expectation value of this condensate signals the 
spontaneous breaking of the $SU(2)$ flavor symmetry down to $U(1)$, with 
two Goldstone pions. Notwithstanding parity is spontaneously broken if 
the first equation in \eqref{orpaao} holds, the vacuum expectation value of the 
flavor diagonal condensate $\langle i\bar\psi\gamma_5\psi\rangle$ vanishes because it 
is also order parameter for a discrete symmetry, composition of parity and 
discrete flavor rotations, which is assumed to be realized \cite{a2}.

Following the lines developed in \cite{pdc}, we wish to write the $p.d.f.$ of the two fermion bilinear order parameters \eqref{orpa}. The motivation to develop 
this formalism was precisely the study of
the vacuum invariance (non-invariance), in quantum theories regularized on a
space time lattice, under symmetry transformations which, as chiral, flavor
or baryon symmetries, involve fermion fields. Notwithstanding that Grassmann 
variables cannot be simulated in a
computer, it was shown in \cite{pdc} that an analysis of spontaneous
symmetry breaking based on the use of the $p.d.f.$ of fermion local
operators can also be done in $QFT$ with fermion degrees of freedom.
The starting point is to choose an order
parameter for the desired symmetry $O(\psi,\bar\psi)$ (typically a
fermion bilinear) and characterize each vacuum state $\alpha$ by the
expectation value $c_\alpha$ of the order parameter in the $\alpha$ state

\be
c_\alpha= \frac{1}{V} \int \langle O(x)\rangle_\alpha d^4 x
\label{exval}
\ee

The $p.d.f.$ $P(c)$ of the order parameter will be given by
\be
P(c)=\sum_\alpha w_\alpha \> \delta (c-c_\alpha)
\label{pdcferm}
\ee

\noindent
which can also be written as \cite{pdc}

\be
P(c)=\left\langle \delta(\frac{1}{V} \int O(x) d^4x -c) \right\rangle
\label{pdcferm2}
\ee

\noindent
the mean value computed with the integration measure of the path
integral formulation of the Quantum Theory.

The Fourier transform $P(q)=\int e^{iqc} P(c) dc$ can be written, for the
model we are interested in, as

$$
P(q) = 
$$
\be
{\frac{\int [dU] [d\bar\psi d\psi]
\exp \{-S_{YM} +\bar\psi W(\kappa)\psi+ {\frac{iq}{V}} \int d^4x
\>\>  O(x)\}}{\int [dU] [d\bar\psi d\psi] \exp
\{-S_{YM}  +\bar\psi W(\kappa)\psi \}} }
\label{pdiqferm}
\ee

In the particular case in which $O$ is a fermion bilinear of $\bar\psi$
and $\psi$

\be
O(x)=\bar\psi(x) \tilde O \psi(x)
\label{bili}
\ee

\noindent
with $\tilde O$ any matrix with Dirac, color and flavor indices, equation
\reff{pdiqferm} becomes

\be
P(q)= {\frac{\int [dU] [d\bar\psi d\psi]
\exp \{-S_{YM} +\bar\psi (W(\kappa)+{\frac{iq}{V}} \tilde O)\psi \}}{\int [dU] [d\bar\psi d\psi] \exp
\{-S_{YM}  +\bar\psi W(\kappa)\psi \}} }
\label{pdiqbili}
\ee

Integrating out the fermion fields in \reff{pdiqbili} one gets

\be
P(q)= {\frac{\int [dU] e^{-S_{YM}} \det(W(\kappa)+{\frac{iq}{V}} \tilde O) }
{\int [dU] e^{-S_{YM}} \det W(\kappa) } }
\label{pdiqbi}
\ee

\noindent
which can also be expressed as the following mean value

\be
P(q)=\left\langle {\frac{\det (W(\kappa)+{\frac{iq}{V}} \tilde O)}{\det W(\kappa)}} \right\rangle
\label{pdiqb}
\ee

\noindent
computed in the effective gauge theory with the integration measure

$$
[dU] e^{-S_{YM}} \det W(\kappa)
$$

\noindent
Notice that zero modes of the Dirac-Wilson operator, which would produce a 
singularity of the operator in \reff{pdiqb}, are suppressed by the fermion 
determinant in the integration measure (zero mode configurations have on the 
other hand vanishing measure).

The particular form expected for the $p.d.f.$ $P(c)$, $P(q)$,
depends on the realization of the corresponding symmetry in the
vacuum. A symmetric vacuum will give

\be
P(c)=\delta(c)
\label{delta}
\ee
$$
P(q)=1
$$

\noindent
whereas, if we have for instance a $U(1)$ symmetry which is
spontaneously broken, the expected values for $P(c)$ and $P(q)$
are \cite{pdc}

$$
P(c)=[ \pi (c_0^2-c^2)^{1/2} ]^{-1} \qquad\qquad  -c_0 < c < c_0
$$
$$
{\rm otherwise} \qquad P(c)=0
$$
\be
P(q)={\frac{1}{2\pi}} \int_{-\pi}^\pi d\theta e^{iqc_0\cos\theta}
\label{pdfu1}
\ee

\noindent
the last being the well known zeroth order Bessel function of the
first kind, $J_0(qc_0)$.

In the simpler case in which a discrete $Z(2)$ symmetry is
spontaneously broken, the expected form is

$$
P(c)= {\frac{1}{2}} \delta(c-c_0) + {\frac{1}{2}} \delta(c+c_0)
$$
\be
P(q)=\cos(q c_0)
\label{pdfz2}
\ee

\noindent
or a sum of symmetric delta functions ($P(c)$) and a sum of
cosines ($P(q)$) if there is an extra vacuum degeneracy.

If we call $P_j(q), P_0(q)$ the p.d.f. of $i\bar\psi\gamma_5\tau_j\psi$ and 
$i\bar\psi\gamma_5\psi$ in momentum space, we have

$$
P_1(q) = P_2(q) = P_3(q) = \left\langle \prod_j \left ( - {\frac{q^2}{V^2
\lambda_j^2}} +1 \right ) \right \rangle
$$
\be
P_0(q) =  \left\langle \prod_j \left ( {\frac{q}{V \lambda_j}} +1 \right )^2
 \right \rangle
\label{pdfmom}
\ee

\noindent
where $V$ is the number of degrees of freedom (including color and Dirac but 
not flavor d.o.f.) and $\lambda_j$ are the real eigenvalues of the 
Hermitian Dirac-Wilson operator $\bar W(\kappa) = \gamma_5 W(\kappa)$. Notice 
that whereas $P_3(q)$ has not a definite sign, $P_0(q)$ is always positive 
definite.

The q-derivatives of $P(q)$ give us the moments of the distribution $P(c)$. 
In particular we have

\be
{ \frac{d^n{P(q)}}{dq^n}} \bigg\rvert_{q=0} = i^n \langle c^n \rangle
\label{qder}
\ee

Since $c$ is order parameter for the symmetries of the lattice action 
and we integrate over all the Gibbs state, the first moment will vanish 
always, independently of the realization of the symmetries. 
The first non-vanishing moment, if the symmetry is spontaneously broken, 
will be the second. Thus, for the particular case

$$
c_0= {\frac{1}{V}} \sum_x i\bar\psi(x)\gamma_5 \psi(x)
$$
\be
c_3= {\frac{1}{V}} \sum_x i\bar\psi(x)\gamma_5 \tau_3 \psi(x)
\label{c0c3}
\ee

we get

$$
\left\langle c_0^2 \right\rangle = 2 \left\langle {\frac{1}{V^2}}
\sum_j {\frac{1}{\lambda_j^2}}\right \rangle  - 4 \left\langle 
\left ( {\frac{1}{V}} \sum_j {\frac{1}{\lambda_j}} \right )^2 \right\rangle
$$
\be
\left\langle c_3^2 \right\rangle = 2 \left\langle {\frac{1}{V^2}}
\sum_j {\frac{1}{\lambda_j^2}}\right \rangle  
\label{vasp}
\ee

In the QCD region flavor symmetry is realized. The p.d.f. of $c_3$ will 
be then $\delta(c_3)$ and $\langle c_3^2\rangle = 0$. We get then

$$
\left\langle c_0^2 \right\rangle = - 4 \left\langle 
\left ( {\frac{1}{V}} \sum_j {\frac{1}{\lambda_j}} \right )^2 \right\rangle, 
$$

\noindent
which should vanish since parity is also realized in this region. Furthermore 
a negative value of $\langle c_0^2\rangle$ would violate Hermiticity of 
$i\bar\psi\gamma_5\psi$. We will come back to this point in the next section.

In the Aoki region \cite{a2} there are vacuum states in which the condensate 
$c_3$ \eqref{c0c3} takes a non-vanishing vacuum expectation value. This implies 
that the p.d.f. $P(c_3)$ will not be $\delta(c_3)$ and therefore $\langle c_3^2\rangle$ 
\eqref{vasp} will not vanish. Indeed expression \eqref{vasp} for $\langle c_3^2\rangle$ seems to 
be consistent with the Banks and Casher formula \cite{BC} which relates 
the spectral density of the Hermitian Dirac-Wilson operator at the origin 
with the vacuum expectation value of $c_3$ \cite{sharpe}.

If, on the other hand, $\langle i\bar\psi\gamma_5\psi\rangle=0$
in one of the Aoki vacua, as conjectured in 
\cite{a2}, $\langle i\bar\psi\gamma_5\psi\rangle=0$ 
in all the other vacua which are connected with the standard 
Aoki vacuum by a parity-flavor transformation, since 
$i\bar\psi\gamma_5\psi$ is invariant under 
flavor transformations and change sign under parity. Therefore if we assume 
that these are all the degenerate vacua, we conclude that 
$P(c_0)= \delta(c_0)$ 
and $\langle c_0^2\rangle=0$, which would imply the following non-trivial relation

\be
\left\langle {\frac{1}{V^2}}
\sum_j {\frac{1}{\lambda_j^2}}\right \rangle  = 2 \left\langle
\left ( \frac{1}{V} \sum_j {\frac{1}{\lambda_j}} \right )^2 \right\rangle
\neq 0
\label{rela}
\ee

\section{QCD with a Twisted Mass Term: the Non Symmetric Case}

In this section we will consider lattice QCD with Wilson fermions with 
the standard action of previous section, plus a source term

\be
\sum_x i m_t \bar\psi(x)\gamma_5\tau_3\bar\psi(x)
\label{source}
\ee

\noindent
that explicitly breaks flavor and parity. The flavor symmetry is thus 
broken from $SU(2)$ to $U(1)$. This is the standard way to 
analyze spontaneous symmetry breaking. First one takes the thermodynamic 
limit and then the vanishing source term limit.

We can calculate again the p.d.f. $\bar P_0(q)$ and $\bar P_3(q)$ of 
$i\bar\psi\gamma_5\psi$ and $i\bar\psi\gamma_5\tau_3\psi$ 
with this modified action. Simple algebra give us the following 
expressions

$$
\bar P_0(q) = \left\langle \prod_j \left ( \frac{\frac{q^2}{V^2} + {\frac{2q}{V} \lambda_j}}{m_t^2+\lambda_j^2} + 1 \right ) \right \rangle
$$
\be
\bar P_3(q) = \left\langle \prod_j \left ( {\frac{{\frac{q^2}{V^2}} + {\frac{2q}{V}} i m_t}{m_t^2+\lambda_j^2}} - 1 \right ) \right \rangle
\label{pdfsou}
\ee

\noindent
where again $\lambda_j$ are the real eigenvalues of the Hermitian 
Dirac-Wilson operator and the mean values are computed now with the 
integration measure of the lattice QCD action modified with the 
symmetry breaking source term.

By taking the q-derivatives at the origin of $\bar P_0(q)$ and $\bar P_3(q)$ 
we obtain

$$
\langle c_0 \rangle = {\frac{2 i}{V}} \left\langle \sum_j {\frac{\lambda_j}{m_t^2+\lambda_j^2}} \right\rangle
$$
$$
\langle c_0^2 \rangle = {\frac{4}{V^2}} \left\langle\sum_j {\frac{\lambda_j^2}{(m_t^2+\lambda_j^2)^2}}
 \right\rangle -
{\frac{2}{V^2}} \left\langle \sum_j {\frac{1}{m_t^2+\lambda_j^2}}\right\rangle -
$$
\be
-4  \left\langle \left ( \frac{1}{V} \sum_j {\frac{\lambda_j}{m_t^2+\lambda_j^2}} \right )^2 \right\rangle
\label{paraor}
\ee

\noindent
and

\be
\langle c_3 \rangle = {\frac{2}{V}} m \left\langle \sum_j {\frac{1}{m_t^2+\lambda_j^2}} \right\rangle
\label{baca}
\ee

Equation \eqref{baca} is well known. If we take the infinite volume limit and then 
the $m\rightarrow 0$ limit we get the Banks and Casher result

\be
\langle c_3 \rangle = 2\pi \rho(0)
\label{dens0}
\ee

\noindent
which relates a non-vanishing spectral mean density of the Hermitian Wilson 
operator at the origin with the spontaneous breaking of parity and flavor 
symmetries.

The first equation in \eqref{paraor} is actually unpleasant since it predicts an 
imaginary number for the vacuum expectation value of a Hermitian operator.
However it is easy to see that $\langle c_0\rangle=0$ because it is order parameter for 
a symmetry of the modified lattice action, the composition 
of parity with discrete flavor rotations around the x or y axis. 

Concerning the second equation in \eqref{paraor}, one can see that the first and 
second contributions to $\langle c_0^2\rangle$ vanish in the infinite volume limit for 
every non-vanishing value of $m$. The third contribution however, which is 
negative, will vanish only if the spectral density of eigenvalues of the 
Hermitian Wilson operator $\rho_U(\lambda)$ for any background gauge field 
$U$ is an even function of $\lambda$. This is actually not true at finite 
values of $V$, and some authors \cite{sharpe,heller} suggest that the 
symmetry of the eigenvalues will be recovered not in the thermodynamic limit, 
but only in the zero lattice spacing or continuum limit. If we take this last 
statement as true, we should conclude:

i. The Aoki phase, which seems not to be connected with the critical 
continuum limit point $(g^2=0, \kappa=1/8)$ \cite{ilg,ster} is unphysical 
since the $\langle c_0^2\rangle$ would be negative in this phase and this result breaks 
Hermiticity.

ii. In the standard QCD phase, where parity and flavor symmetries are 
realized in the vacuum, we should have however negative values for the 
vacuum expectation value of the square of the Hermitian operator 
$i\bar\psi\gamma_5\psi$, except 
very near to the continuum limit. Since this operator is related to the 
$\eta$-meson, one can expect in such a case important finite lattice spacing 
effects in the numerical determinations of the $\eta$-meson mass.

This is the first of the two possible scenarios mentioned in the first section
of this article. In the next section we will assume a symmetric spectral
density of eigenvalues of the
Hermitian Wilson operator $\rho_U(\lambda)$ for any background gauge field
$U$ in the thermodynamic limit, and will derive the second scenario.

\section{QCD with a Twisted Mass Term: Symmetric Spectral Density of 
Eigenvalues}

In this section we will assume that the spectral
density of eigenvalues of the
Hermitian Wilson operator $\rho_U(\lambda)$ for any background gauge field
$U$ is an even function of $\lambda$. In such a case equation \eqref{paraor} will
give a vanishing value for $\langle c_0^2\rangle$ at any value of $m_t$

\be
\langle c_0^2 \rangle = 0 
\label{clust}
\ee

\noindent
Therefore the $p.d.f.$ of $c_0$ is $\delta(c_0)$ and 

\be
\langle i \bar\psi\gamma_5\psi \rangle = 0
\label{inva}
\ee

\noindent
for any value of $m_t$, and also in the $m_t\rightarrow 0$ limit. Thus
we can confirm that under the assumed condition,
$\langle i\bar\psi\gamma_5\psi\rangle=0$ in the Aoki vacuum selected by the external
source \eqref{source}, as stated in \cite{a2}; but since $i\bar\psi\gamma_5\psi$
is flavor invariant and change sign under parity, we can conclude that
$\langle i\bar\psi\gamma_5\psi\rangle=0$, not only in the vacuum selected by the
external source \eqref{source}, but also in all the Aoki vacua which can be obtained
from the previous one by parity-flavor transformations. In order to see
the fact that, if there is an Aoki phase with parity-flavor symmetry spontaneously
broken, the previous vacua are not all the possible vacua, we will assume 
that is false and will get a contradiction. 

If all the vacua are the one 
selected by the twisted mass term and those obtained from it by 
parity-flavor transformations, the spectral density of the Hermitian Wilson 
operator will be always an even function of $\lambda$, since the eigenvalues 
of this operator change sign under parity and are invariant under flavor 
transformations. Then the symmetry of 
$\rho_U(\lambda)$ will be realized also at $m_t=0$ in the 
Gibbs state. Now let us come
back to expression \eqref{vasp} which give us the vacuum expectation values
of the square of $i\bar\psi\gamma_5\psi$ and $i\bar\psi\gamma_5\tau_3\psi$
as a function of the spectrum of the Hermitian Wilson operator, but
averaged over all the Gibbs state (without the external symmetry breaking
source \eqref{source}). By subtracting the two equations in \eqref{vasp} we get

\be
\langle c_3^2 \rangle - \langle c_0^2 \rangle = 
4 \left\langle \left ( \frac{1}{V} \sum_j {\frac{1}{\lambda_j}} \right )^2
\right \rangle
\label{diffe}
\ee

This equation would \emph{naively} vanish, if the spectral density of eigenvalues of the Hermitian Wilson operator
were an even function of $\lambda$. Therefore we would reach the following conclusion for the Gibbs state

\be
\langle c_3^2 \rangle = \langle c_0^2 \rangle
\label{gibsta}
\ee

Nevertheless, S. Sharpe put into evidence in a private communication (developed deeply in \cite{sharpe2}) an aspect that we, somewhat, overlooked: A sub-leading contribution to the spectral density may affect \eqref{gibsta} in the Gibbs state ($\epsilon$-regime, in $\chi$PT terminology), in such a way that, not only $\langle c_0^2 \rangle$, but every even moment of $i\bar\psi\gamma_5\psi$ would vanish, restoring the standard Aoki picture. The thesis of Sharpe, although possible, would enforce an infinite series of \emph{sum rules} to be complied, similar to those found by Leutwyler and Smilga in the continuum \cite{Leut}. We agree that such a possibility is open, at
least from a purely mathematical point of view: In fact sub-leading
contributions to the spectral density may exist, and conspire to 
enforce the vanishing of all the even moments of the $p.d.f.$ of 
$i\bar\psi\gamma_5\psi$. However 
we believe such possibility not to be very realistic, and indeed we have 
physical arguments, which will be the basis for subsequent work on the 
topic, suggesting that the Aoki scenario is incomplete. Therefore, we will reasonably assume in the following \eqref{gibsta} to be true, in the case of a symmetric $\rho_U(\lambda)$, assumption that leads us to the conclusion that the \emph{Chiral Perturbation Theory may be incomplete}, for the new vacua derived from \eqref{gibsta} are not predicted in $\chi$PT.

If as conjectured by Aoki and verified by numerical simulations, a
phase with a non-vanishing vacuum expectation value of
$i\bar\psi\gamma_5\tau_3\psi$ does exists,
the mean value in the Gibbs state $\langle c_3^2\rangle$ inside this phase will be non-zero.
Then equation \eqref{gibsta} tell us that also $\langle c_0^2\rangle$ will be non-zero inside
this phase. But since in the Aoki vacua $\langle c_0^2\rangle=0$, this is in contradiction 
with the assumption that the Aoki vacua are all possible vacua.
This is the second possible scenario mentioned in the Introduction of this 
article.

\section{The New Vacua}

To understand the physical properties of these new vacuum states we will 
assume, inspired by the numerical results reported in the next section, 
that the spectral density of eigenvalues $\rho_U(\lambda)$ is an even 
function of $\lambda$ in the Gibbs state of the Aoki region ($m_t=0$). Then
equation \eqref{gibsta} holds (taking into account the aforementioned discussion raised by S. Sharpe), and hence the $p.d.f.$ of the flavor singlet
$i\bar\psi\gamma_5\psi$ order parameter can not
be $\delta(c_0)$ inside the Aoki phase, and therefore new vacuum states
characterized by a non-vanishing vacuum expectation value of
$i\bar\psi\gamma_5\psi$ should appear.
These new vacua can not be connected, by mean of parity-flavor
transformations, to the Aoki vacua, as previously discussed.

In order to better characterize these new vacua, we have added to the lattice
QCD action the source term

\be
i m_t \bar\psi\gamma_5\tau_3\psi + i \theta\bar\psi\gamma_5\psi 
\label{2sour}
\ee

\noindent
which breaks more symmetries than \eqref{source}, but still preserves the $U(1)$
subgroup of the $SU(2)$ flavor. By computing again the first moment of the
p.d.f. of $i\bar\psi\gamma_5\psi$ and $i\bar\psi\gamma_5\tau_3\psi$ and
taking into account that the mean value of the first of these operators is
an odd function of $\theta$ whereas the second one is an even function of
$\theta$, we get

$$
\langle i\bar\psi\gamma_5\psi \rangle = -{\frac{2\theta}{V}} \left\langle
\sum_j {\frac{-\lambda_j^2+m_t^2-\theta^2} {(\lambda_j^2+m_t^2-\theta^2)^2
+4\theta^2\lambda_j^2}}\right\rangle 
$$
\be
\langle i\bar\psi\gamma_5\tau_3\psi \rangle = {\frac{2m_t}{V}} \left\langle
\sum_j {\frac{\lambda_j^2+m_t^2-\theta^2}{(\lambda_j^2+m_t^2-\theta^2)^2
+4\theta^2\lambda_j^2}}\right\rangle
\label{fimo}
\ee

\noindent
where $\lambda_j$ are again the eigenvalues of the Hermitian Wilson operator
and the mean values are computed using the full integration measure of
lattice QCD with the extra external source \eqref{2sour}. This
integration measure is not positive definite due to the presence of the
$i\bar\psi\gamma_5\psi$ term in the action.

By choosing $\theta=rm_t$ in the action and taking the thermodynamic limit
we get for the two order parameters the following expressions

$$
\langle i\bar\psi\gamma_5\psi \rangle = \int {\frac{2rm_t\lambda^2-2rm_t^3
(1-r^2)}{\left( m_t^2(1-r^2)+\lambda^2\right)^2+4r^2m_t^2\lambda^2}}
\rho(\lambda) d\lambda
$$
\be
\langle i\bar\psi\gamma_5\tau_3\psi \rangle = \int {\frac{2m_t^3(1-r^2)+
2m_t\lambda^2}{\left( m_t^2(1-r^2)+\lambda^2\right)^2
+4r^2m_t^2\lambda^2}}
\rho(\lambda) d\lambda
\label{inte}
\ee

\noindent
where $\rho(\lambda)$ is the mean spectral density of the Hermitian
Wilson operator averaged with the full integration measure.

Taking now the $m_t\rightarrow 0$ limit i.e., approaching the vanishing 
external
source \eqref{2sour} point in the $\theta, m_t$ plane on a line crossing the origin and
with slope r, we obtain

$$
\langle i\bar\psi\gamma_5\psi \rangle = 2\rho(0)\int_{-\infty}^{+\infty} 
{\frac{rt^2-r(1-r^2)}{\left( 1-r^2+t^2\right)^2+4r^2t^2}}dt
$$
\be
\langle i\bar\psi\gamma_5\tau_3\psi \rangle = 2\rho(0)\int_{-\infty}^{+\infty} 
{\frac{1-r^2+t^2}{\left( 1-r^2+t^2\right)^2+4r^2t^2}}dt
\label{intem0}
\ee

In the particular case of $r=0$ ($\theta=0$) we get the Banks and Casher
formula

\be
\langle i\bar\psi\gamma_5\tau_3\psi \rangle = 2\pi\rho(0)
\label{baca2}
\ee

We see how, if $\rho(0)$ does not vanish, we can get many vacua
characterized by a non-vanishing value of the two order parameters
$i\bar\psi\gamma_5\psi$ and $i\bar\psi\gamma_5\tau_3\psi$. We
should point out that the value of $\rho(0)$ could depend on the slope
$r$ of the straight line along which we approach the origin in the
$\theta, m$ plane, and therefore, even if results of numerical simulations 
suggest that
$\rho(0)\ne 0$ when we approach the origin along the line of vanishing slope,
this does not guarantee that the same holds for other slopes. However the
discussion in the first half of this section tell us that if $\rho(0)\ne 0$
at $r=0$, $\rho(0)$ should be non-vanishing for other values of $r$.

\section{Quenched Numerical Simulations}

In order to distinguish what of the two possible scenarios derived in the
previous sections is realized, we have performed quenched simulations of
lattice QCD with Wilson fermions in $4^4, 6^4$ and $8^4$ lattices.
We have generated an ensemble of well uncorrelated configurations
for each volume and then a complete diagonalization of the Hermitian
Wilson matrix, for each configuration, gives us the respective eigenvalues.
We measured the volume dependence of the asymmetries in the eigenvalue
distribution of the Hermitian Wilson operator, both inside and outside
the Aoki phase.

We want to notice that because of kinematic reasons (properties of the Dirac
matrices), the trace of all odd p-powers of the Hermitian Wilson operator
$\bar W(\kappa) = \gamma_5 W(\kappa)$
vanish until $p=7$, this included. This means that the asymmetries in the
eigenvalue distribution of
$\bar W(\kappa)$ start to manifest with a non-vanishing
value of the ninth moment of the distribution. We have found that these
asymmetries, even if small, are clearly visible in the numerical simulations.

In Figs. from 2 to 6 we plot the quenched mean value

\be
A(\beta,\kappa,m_t) = \left \langle \left ( \frac{1}{V} \sum_j { \frac{\lambda_j}{m_t^2+\lambda_j^2}} \right )^2 \right \rangle_Q
\label{qmv}
\ee
multiplied by the volume for three different volumes, in order to see the scaling of the asymmetries in the eigenvalue distribution of $\bar W(\kappa)$. As previously discussed, $A(\beta,\kappa,m_t)$ give us a quantitative measure of these asymmetries. We have added an extra $V$ factor to make the plots for the three different volumes distinguishable. Since we found that the value of $A(\beta,\kappa,m_t)$ decreased as the volume increased, the plots of the larger volumes were negligible with respect to the plot of the smaller volume $4^4$. Multiplying by $V$ all the plots are of the same magnitude order.

The $m_t$
term in the denominator of \eqref{qmv} acts also as a regulator in the quenched
approximation, where configurations with zero or near-zero modes are not
suppressed by the fermion determinant. This is very likely the origin of the
large fluctuations observed in the numerical measurements of \eqref{qmv} near $m_t=0$
in the quenched case. That is why our plots are cut below $m_t=0.05$; in the physical phase, this cutoff is not really  needed, but in the Aoki phase it is more likely to find zero modes which spoil the distribution.

Figs. 2 and 3 contain our numerical results in
$4^4, 6^4$ and $8^4$ lattices at $\beta=0.001, \kappa=0.17$ and
$\beta=5.0, \kappa=0.15$. These first two points are
outside the Aoki phase, the first one in the strong coupling region. The second one intends to be a point where typically QCD simulations are performed.

\begin{figure}[h]
\resizebox{8.3 cm}{!}{\includegraphics{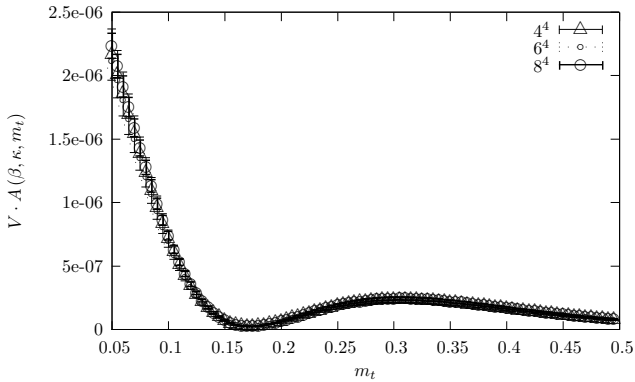}}
\caption{Point outside of the Aoki phase ($\beta = 0.001$, $\kappa = 0.17$) and in the strong coupling regime. The superposition of plots clearly states that the asymmetry of the eigenvalue distribution decreases as $\frac{1}{V}$. Statistics: 240 configurations ($4^4$), 2998 conf. ($6^4$) and 806 conf. ($8^4$)}
\end{figure}

\begin{figure}[h]
\resizebox{8.3 cm}{!}{\includegraphics{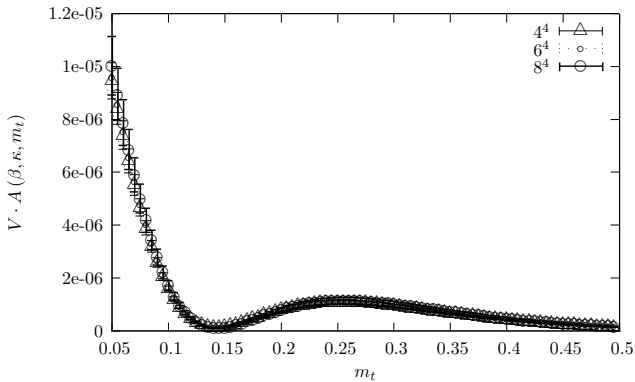}}
\caption{Another point outside of the Aoki phase ($\beta = 5.0$, $\kappa = 0.15$) in a region in which QCD simulations are commonly performed. The conclusion is the same as in Fig. 2. Statistics: 400 conf. ($4^4$), 900 conf. ($6^4$) and 200 conf. ($8^4$)}
\end{figure}

Figs. 4, 5 and 6 represent our numerical results in
$4^4, 6^4$ and $8^4$ lattices at $\beta=0.001, \kappa=0.30$, $\beta=3.0, \kappa=0.30$ and
$\beta=4.0, \kappa=0.24$. These points are well inside the Aoki phase, and the structure of the distribution is different from the structure observed in the previous plots of the physical phase. Nevertheless, the qualitative behaviour as the volume increases is the same.

\begin{figure}[h]
\resizebox{8.3 cm}{!}{\includegraphics{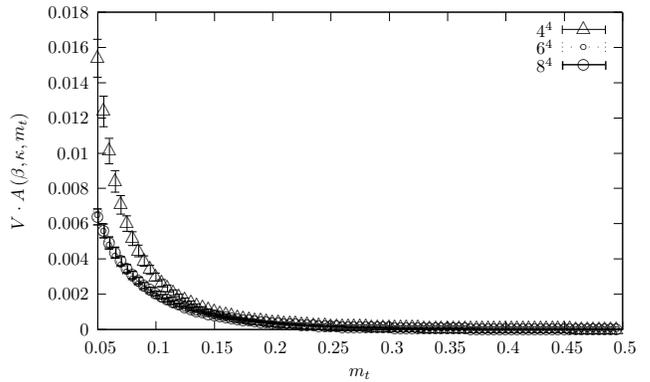}}
\caption{Point inside the Aoki phase ($\beta = 0.001$, $\kappa = 0.30$) and in the strong coupling regime. Although there is no clear superposition of plots, it is evident that the asymmetry goes to zero as the volume increases. Statistics: 368 conf. ($4^4$), 1579 conf. ($6^4$) and 490 conf. ($8^4$)}
\end{figure}

\begin{figure}[h]
\resizebox{8.3 cm}{!}{\includegraphics{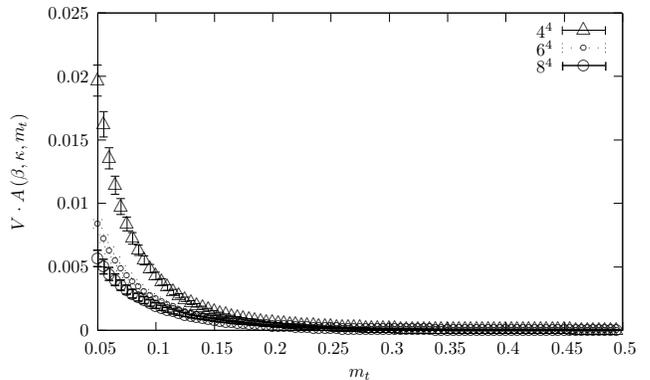}}
\caption{Point inside the Aoki phase ($\beta = 3.0$, $\kappa = 0.30$). The asymmetry disappears as the volume increases. Statistics: 400 conf. ($4^4$), 1174 conf. ($6^4$) and 107 conf. ($8^4$)}
\end{figure}

\begin{figure}[h]
\resizebox{8.3 cm}{!}{\includegraphics{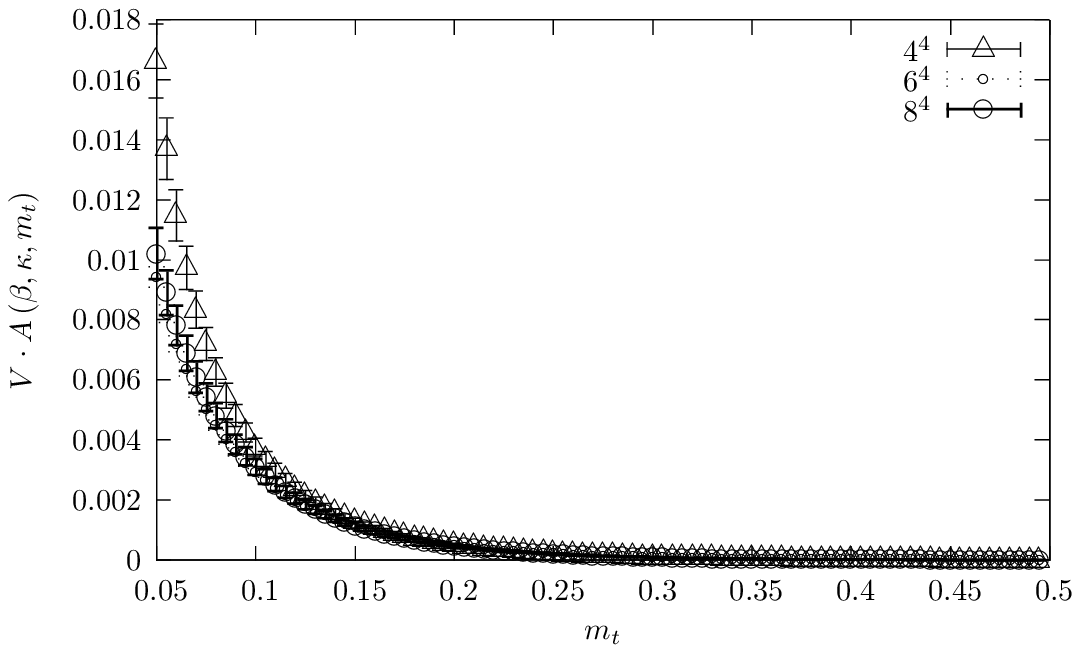}}
\caption{Point inside the Aoki phase ($\beta = 4.0$, $\kappa = 0.24$). Same conclusions as in the other Aoki plots. Statistics: 398 conf. ($4^4$), 1539 conf. ($6^4$) and 247 conf. ($8^4$)}
\end{figure}

We observe large fluctuations in the plotted quantity
near $m_t=0$, specially inside the Aoki phase. However the behavior with the lattice volume may suggests a
vanishing value of $A(\beta,\kappa,m_t)$ in the infinite volume limit in both regions, inside and outside the Aoki phase.
If this is actually the case in the unquenched model, the second scenario
discussed in this article would eventually be realized.

\section{Conclusions}

We have analyzed the vacuum structure of lattice QCD with two degenerate
Wilson flavors at non-zero lattice spacing, with the help of the probability
distribution function of the parity-flavor fermion bilinear order parameters.
From this analysis two possible scenarios emerge.

In the first scenario we assume a spectral
density $\rho_U(\lambda,\kappa)$ of the Hermitian
Dirac-Wilson operator, in a fixed
background gauge field $U$, not symmetric in $\lambda$. This property is
realized at finite $V$ for the single gauge configurations, even if a
symmetric distribution of eigenvalues is recovered if we average over
parity conjugate configurations. We find that under
such an assumption, Hermiticity of
$i\bar\psi_u\gamma_5\psi_u+i\bar\psi_d\gamma_5\psi_d$ will
be violated at finite $\beta$. This lost of Hermiticity for the pseudoscalar
flavor singlet operator suggests that a reliable determination of the $\eta$
mass would require to be near enough the continuum limit where the symmetry
of the spectral density $\rho_U(\lambda,\kappa)$ should be recovered.
Furthermore assuming that the Aoki phase ends at finite $\beta$, the
violation of Hermiticity obtained in this scenario implies the lost
of any physical interpretation of this phase in terms of particle excitations.

In the second scenario a symmetric spectral
density $\rho_U(\lambda,\kappa)$ of the Hermitian Dirac-Wilson operator
in the infinite volume limit is assumed, and then we show that the
existence of the Aoki phase implies also the appearance of other phases, in
the same parameters region, which can be
characterized by a non-vanishing vacuum expectation value of
$i\bar\psi_u\gamma_5\psi_u+i\bar\psi_d\gamma_5\psi_d$,
and with vacuum states that can not be connected with the Aoki vacua by parity-flavor symmetry
transformations. These phases, however, are not related to those mentioned in \cite{GM}, for we keep the twisted mass parameter $m_5$ equal to zero, whereas the phases studied by G. M\"unster appear at large $m_5$.

Sharpe and Singleton \cite{sharpe,shasi} performed an analysis of lattice QCD with two flavors of Wilson 
fermions near the continuum limit, by mean of the chiral effective Lagrangians. In their analysis, they 
found essentially two possible realizations, depending on the sign of a coefficient $c_2$, which appears 
in the potential energy, expanded up to second order in the quark mass term. If $c_2$ is positive, a 
phase with spontaneous flavor symmetry breaking and an Aoki vacuum can be identified. 
If, on the contrary, $c_2<0$, flavor symmetry is realized in the vacuum and a first order 
transition should appear. Either case may be realized in different regions of parameter space.
Indeed numerical simulations with dynamical fermions, performed at small lattice spacing \cite{jansen},
give evidences of metastability that can be related with the existence of a first order phase 
transition and hence $c_2<0$, while the Aoki phase found at smaller $\beta$ values \cite{ilg} supports
$c_2>0$. We want to emphasize that the new vacua we find are coexisting with the standard Aoki vacuum. 
These new vacua do exist if, and only if, the Aoki vacuum exists. But these new vacua are not explained in any way in the chiral effective 
Lagrangian approximation of Sharpe and Singleton. On the other hand, in order to recover the standard Aoki picture via $\chi$PT, an infinite set of sum rules for the eigenvalues of the Hermitian Wilson operator must be imposed. As we stated previously, this possibility does not appeal us, in the sense that it seems unphysical (we would call it a `mathematical' possibility). Nevertheless we have not definite proof of the new vacua.

From our position, these conclusions lead us to cast a doubt into the completeness of the Chiral Perturbation theory. In any case we believe that, in order to definitely clarify this issue, 
a careful investigation of the spectral properties of Hermitian Dirac-Wilson
operator for actual gauge field configuration in the full unquenched
theory is mandatory.

In order to distinguish what of the two possible scenarios, as mentioned at the beginning 
of this section, is realized, we performed quenched simulations of
lattice QCD with Wilson fermions in $4^4, 6^4$ and $8^4$ lattices, and
measured the volume dependence of the asymmetries in the eigenvalue
distribution of the Hermitian Wilson operator, both inside and outside
the Aoki phase. To measure this asymmetries we diagonalized exactly
the Hermitian Wilson operator for all the gauge configurations generated
with the quenched measure, and measured the quenched average
$A(\beta,\kappa,m)$ \eqref{qmv}. Due to the fact that configurations with zero or
near-zero modes are not suppressed by the fermion determinant in the quenched
case, this quantity fluctuates violently near $m=0$, specially for the
points in the Aoki phase. Notwithstanding that,
the observed behavior with the lattice volume seems to suggest a
vanishing value of $A(\beta,\kappa,m)$ in the infinite volume limit, both
inside and outside the Aoki phase.
If this were actually the case in the unquenched model, the second scenario
discussed in this article would be realized. However a verification of
these results in the unquenched case would be very relevant in order to
discard the first scenario.

\acknowledgments

It is a pleasure to thank Fabrizio Palumbo for useful discussions. We also thank Steve Sharpe for his sharp comments and remarks.
This work has been partially supported by an INFN-MEC collaboration, CICYT (grant FPA2006-02315) 
and DGIID-DGA (grant2007-E24/2).

\end{document}